\documentclass[useAMS,usenatbib]{mnras}

\usepackage{amssymb}
\usepackage{amsmath}
\usepackage{relsize}
\usepackage{graphicx}
\usepackage{hyperref}
\usepackage{mathptmx}
\usepackage{color}

\newcommand{\be}{\begin{equation}}
\newcommand{\ee}{\end{equation}}

\makeatletter
\newcommand*{\rom}[1]{\expandafter\@slowromancap\romannumeral #1@}
\makeatother

\newcommand{\wfcode}{\textsc{WarmAndFuzzy}}
\newcommand{\hmcode}{\textsc{HMcode}}

\newcommand{\oneh}{\ensuremath{\Delta^2_{\rm 1H}(k)}}
\newcommand{\twoh}{\ensuremath{\Delta^2_{\rm 2H}(k)}}
\newcommand{\dcrit}{\ensuremath{\delta_{\rm crit}}}

\begin{document}
\newcommand{\tkDM}[1]{\textcolor{red}{#1}}  
\newcommand{\tkBB}[1]{\textcolor{blue}{#1}}  
\newcommand{\tkRW}[1]{\textcolor{green}{#1}}  
\newcommand{\unit}[1]{\ensuremath{\, \mathrm{#1}}}

\newcommand{\Lya}{Lyman-$\alpha$~}

\title{\textsc{WarmAndFuzzy}: the halo model beyond CDM}
\author[D. J. E. Marsh] {David~J.~E.~Marsh$^1$\\
$^1$King's College London, Strand, London, WC2R 2LS, United Kingdom
}
\date{\today}
\maketitle


\begin{abstract}

Cold dark matter (CDM) is a well established paradigm to describe cosmological structure formation, and works extraordinarily well on large, linear, scales. Progressing further in dark matter physics requires being able to understand structure formation in the non-linear regime, both for CDM and its alternatives. This short note describes a calculation, and accompanying code, \textsc{WarmAndFuzzy}, incorporating the popular models of warm and fuzzy dark matter (WDM and FDM) into the standard halo model to compute the non-linear matter power spectrum. The FDM halo model power spectrum has not been computed before. The FDM implementation models ultralight axions and other scalar fields with $m_a\approx 10^{-22}\text{ eV}$. The WDM implementation models thermal WDM with mass $m_X\approx 1\text{ keV}$. The halo model shows that differences between WDM, FDM, and CDM survive at low redshifts in the quasi-linear and fully non-linear regimes. The code uses analytic transfer functions for the linear power spectrum, modified collapse barriers in the halo mass function, and a modified concentration-mass relationship for the halo density profiles. Modified halo density profiles (for example, cores) are not included, but are under development. Cores are expected to have very minor effects on the power spectrum on observable scales. Applications of this code to the Lyman-$\alpha$ forest flux power spectrum and the cosmic microwave background lensing power spectrum will be discussed in companion papers. \textsc{WarmAndFuzzy} is available online at \url{https://github.com/DoddyPhysics/HMcode}, where collaboration in development is welcomed.


\end{abstract}


\section{Introduction}
\label{sec:intro} 

Precision cosmology is all about extracting as much information as possible from cosmological observables with the dual aims of ever more accurate measurements of the parameters of the standard cosmological model, and searching for new parameters that may give clues to the fundamental theory that underpins cosmology. 

On the largest scales, cosmological perturbation theory \citep[e.g.][]{1995ApJ...455....7M,2002PhR...367....1B} provides an accurate description of the growth of structure. It is the foundation of precision parameter estimation from the cosmic microwave background (CMB) and galaxy surveys \citep[e.g.][]{2015arXiv150201589P,2014MNRAS.444.3501B} using publicly available Boltzmann codes such as \textsc{camb}~\citep{camb}. 

In the fight against cosmic variance, we can only win by collecting more modes, which means measuring observables on ever smaller scales, where we encounter non-linearities in the density fields. In the non-linear regime one typically has to rely on computationally expensive approximations such as $N$-body simulations, e.g. \textsc{gadget} \citep{2005MNRAS.364.1105S}, or field-based alternatives \citep{Widrow&Kaiser1993,2014NatPh..10..496S}. Computing these non-linearities is already important for cosmological parameter estimation from galaxy weak lensing~\citep[e.g.][]{2013MNRAS.432.2433H,2015MNRAS.451.2877M}. It will become more important as CMB lensing moves to higher multipoles with Stage-IV CMB experiments~\citep[e.g.][]{2014ApJ...788..138W}, and will be vitally important to the success of future galaxy weak lensing surveys such as \emph{Euclid}~\citep[e.g.][]{2013LRR....16....6A}.

The so-called ``halo model'' of large scale structure \citep{2000MNRAS.318..203S,2000MNRAS.318.1144P,2002PhR...372....1C} is an important tool for computing the matter power spectrum, $\Delta^2_m(k)$, beyond perturbation theory. Though somewhat heuristic, it is based on well-founded physical principles and provides qualitative and quantitative agreement with the results of $N$-body simulations~\citep[e.g.][hereafter M15]{2015MNRAS.454.1958M}. Being semi-analytic, it is a computationally inexpensive method to explore the effects of non-linearities on the cosmological clustering of matter. The halo model can also be used to analytically compute things beyond the power spectrum, such as higher point statistics and halo bias, though I will not pursue such calculations here.

In this short note I explain a new code called \wfcode~that calculates halo model power spectra in models of dark matter (DM) beyond the standard cold DM (CDM). \wfcode~uses \hmcode~by M15 \citep[see also][]{2015ascl.soft08001M} as a base for the halo model calculation, but by default does \emph{not} use the modified fit improvements of M15 as these have not been tested for WDM or FDM.\footnote{Inclusion of massive neutrinos and modified gravity in \textsc{HMcode} was discussed in \cite{2016MNRAS.459.1468M}.} The two models I implement are warm DM \citep[WDM, e.g.][]{2001ApJ...556...93B} and ``fuzzy'' DM \citep[FDM, e.g.][]{2000PhRvL..85.1158H}. The WDM models I implement correspond to thermal relics, such as gravitinos~\citep[e.g.][]{1982PhRvL..48..223P,1982PhRvL..48.1636B} with mass $m_X\approx 1\text{ keV}$.\footnote{Sterile neutrinos~\citep[e.g.][]{1994PhRvL..72...17D} are also a plausible WDM candidate, but in this initial exploration I do not treat them. They are qualitatively similar to thermal relics, but with quantitative differences \cite[e.g.][]{2016MNRAS.459.1489B}. With the appropriate transfer function, they can be treated in \wfcode~the same as thermal relics.} The FDM models I implement correspond to scalar fields non-thermally produced by vacuum realignment with mass $m_a\approx 10^{-22}\text{ eV}$, which could be axions~\citep[e.g.][]{axiverse,2016PhRvD..93b5027K,2015arXiv151007633M} or other scalars~\citep[e.g.][]{1995PhRvL..75.2077F,2000CQGra..17.1707M,2014ASSP...38..107S}. The FDM model assumes that the potential is well-approximated by $m_a^2\phi^2$, i.e. applies in the limit that the self-interactions are negligible in their effects on the transfer function and the halo mass function (HMF).\footnote{For an axion with the canonical cosine potential, this is true for models with large decay constants. Large decay constants are typically necessary to obtain the correct relic abundance at the ultralight masses considered, without recourse to additional tuning or production mechanisms.} FDM self-interactions are discussed further at the end of this paper.

\wfcode~uses modifications to the linear transfer function (including power suppression), the collapse barrier in the \cite{1974ApJ...187..425P} formalism for the HMF (including a halo mass dependence), and to the halo concentration parameter of \cite{1997ApJ...490..493N} (NFW) halo density profiles. The ingredients of the halo model for FDM were developed in \cite{2014MNRAS.437.2652M}, but they have so far not been applied to an efficient calculation of $\Delta_m^2(k)$. \wfcode~is the first attempt to calculate the halo model power for FDM, and this is the primary purpose of the code. The halo model has already been developed for WDM~\citep{2011PhRvD..84f3507S,2011arXiv1109.6291D,2012MNRAS.424..684S}. The implementation in \wfcode~is somewhat simpler than these models, but it is expected to give correct results over a wide range of scales, and incorporates some different physical principles (modified barrier over fits to simulation in the HMF). I have included WDM essentially as an ``added bonus'' thanks to its similarity to FDM. I discuss the limitations of my modelling of FDM and WDM later and throughout this paper.

Section~\ref{sec:transfers} presents the modified linear transfer functions used, Section~\ref{sec:barriers} presents the modified collapse barriers, Section~\ref{sec:concentrations} presents the modified concentration-mass relationship, and Section~\ref{sec:examples} presents the results for the non-linear power. I discuss some limitations and the expected accuracy of \wfcode~in Section~\ref{sec:discussion}, and then conclude. In Appendix~\ref{sec:halomodel} I give a brief description of the halo model as implemented in M15, used in both \hmcode~and \wfcode. I discuss the minor technicalities of \wfcode~in Appendix~\ref{sec:code}. 

\section{Modifying the halo model}
\label{sec:main} 

The basic halo model used in \hmcode~and \wfcode~is described in Appendix~\ref{sec:halomodel}, where I also define much of my notation. Here, I only describe the new aspects added for FDM and WDM. 

I compute the HMF for FDM and WDM using the halo model with a modified linear transfer function, and a mass-dependent modified collapse barrier in the \cite{1974ApJ...187..425P} formalism for the HMF. This differs from the HMF fit provided for FDM by \cite{2016ApJ...818...89S}, which includes the effects of the linear transfer function, and the removal of ``spurious structure'' \citep{2007MNRAS.380...93W} but does include the modified barrier effects (though it is noted that such effects become important on similar scales).\footnote{The scalar field based code of \cite{2014NatPh..10..496S} does include such effects, but has not yet been employed on large enough simulations to estimate the HMF.} My approach differs from that used by many authors studying WDM \citep[e.g][]{2011PhRvD..84f3507S}, who also use by-hand cut-offs fit to simulations based on removal of spurious structure, and not the modified barrier found including thermal velocities~\citep{2001ApJ...558..482B}. 

\subsection{The linear theory power spectrum}
\label{sec:transfers}

By default \hmcode~models the linear theory matter power spectrum for CDM (+baryons) using the analytic fit of \cite{1998ApJ...496..605E}. \wfcode~supplements this with two analytic fits for the relative effects of FDM~\citep{2000PhRvL..85.1158H} and WDM~\citep{2001ApJ...556...93B}, such that the linear power spectrum is given by:
\be
P_X(k)=T_X^2(k)P_{\rm CDM}(k) \, ,
\ee
where I have suppressed redshift dependence of $P(k)$, and the transfer functions, $T_X(k)$, are assumed to be redshift-independent. The transfer functions are given by:
\begin{align}
T_{\rm W}(k) &=[1+(\alpha k)^{2\mu}]^{-5/\mu} \, , \label{eqn:wdm_transfer}\\
T_{\rm F}(k) &= \frac{\cos x_J^3(k)}{1+x_J^8(k)} \, . \label{eqn:fcdm_transfer}
\end{align}
The fitting parameters are
\begin{align}
\mu &= 1.12 \, , \\
\alpha &= 0.074 \left(\frac{m_X}{\text{keV}}\right)^{-1.15}\left(\frac{0.7}{h}\right) \text{ Mpc}\, , \\
x_J(k)&=1.61 \left(\frac{m_a}{10^{-22}\text{ eV}} \right)^{1/18}\frac{k}{k_{J,{\rm eq}}} \, , \\
k_{J,{\rm eq}}&=9 \left(\frac{m_a}{10^{-22}\text{ eV}} \right)^{1/2}\text{ Mpc}^{-1} \, .
\end{align}

These transfer functions cause the linear power spectrum in FDM and WDM to be suppressed relative to CDM below a characteristic value of $k$. For WDM this is the free-streaming wavenumber, $k_{\rm fs}$, caused by thermal velocities. For FDM this is the scalar field Jeans wavenumber, $k_J$, caused by the gradient energy in the Klein-Gordon equation, which manifests as an effective pressure in the fluid equations~\citep{khlopov_scalar}. The FDM transfer function has a sharper cut-off than the WDM transfer function, and displays acoustic oscillations on small scales below $k_{J,{\rm eq}}$.  Both the Jeans wavenumber and the free-streaming wavenumber increase for increasing particle mass, such that models of FDM with $m_a\gg 10^{-22}\text{ eV}$ and WDM with $m_X\gg 1\text{ keV}$ look increasingly like CDM on astrophysically observable scales. 

These transfer functions apply only for DM models composed \emph{entirely} of FDM \emph{or} WDM. In this case the FDM transfer function matches the full Boltzmann code calculation of \cite{2015PhRvD..91j3512H} reasonably well, reproducing the main differences to WDM.\footnote{Mixed DM models are not supported by \wfcode, though further simple modifications could approximate them. See e.g. \cite{2006PhLB..642..192A,2010PhRvD..82j3528M} for possible transfer function fits, or use an input numerical spectrum following e.g. \cite{2015PhRvD..91j3512H,2015arXiv151108195U}. However, note that for mixed dark matter the modified collapse barriers in \wfcode, discussed below, will have to be further modified or also entered numerically, as in \cite{2014MNRAS.437.2652M,2015MNRAS.450..209B}.} 

\subsection{Modified collapse barriers}
\label{sec:barriers}

In spherical collapse of CDM, the critical overdensity barrier for collapse, $\dcrit$, can be derived analytically, and is given by $\dcrit^0=(3/20)\times (12\pi)^{2/3}\approx 1.686$ \citep[e.g.][]{1993ppc..book.....P}. In the \cite{1974ApJ...187..425P} formalism used in the halo model, one can trivially absorb the redshift dependence of the power spectrum into the barrier:
\be
\dcrit (z) = \frac{\dcrit^0}{D(z)} \, ,
\ee
where $D(z)$ is the linear growth function, normalised to unity at $z=0$. The barrier is effectively larger at early times, which accounts for the smaller density perturbations and the use of the power spectrum variance, $\sigma^2(M)$, fixed at $z=0$. 

\wfcode~employs a \emph{mass-dependent} critical overdensity for collapse, implemented as
\be
\dcrit (M,z) =\mathcal{G}_X(M) \frac{\delta_{\rm crit}^0}{D(z)} \, ,
\label{eqn:mass_dependent_barrier}
\ee
where I have suppressed the particle mass, and possible redshift, dependence in $\mathcal{G}_X(M)$. 

Eq.~\eqref{eqn:hmf_def_standard} defines the HMF in the case of a mass-independent barrier. For a mass-dependent barrier, one must be more careful: the differential refers explicitly to the variance, which is compared to the barrier size. Thus the mass function in this case is defined as \citep[see, e.g.][]{2004ApJ...609..474B}
\be
\frac{dn}{dM} dM = -\frac{\bar{\rho}}{M}f(\nu) \frac{d\sigma^2}{\sigma^2} \, ,
\label{eqn:hmf_def_mass_barrier}
\ee
where $\nu = \dcrit(M,z)/\sigma(M)$, and $f(\nu)$ is the \cite{1999MNRAS.308..119S} function as defined in Appendix~\ref{sec:halomodel}. The important change in the case of the mass dependent barrier is that $\sigma$ and $\dcrit$ are now independent variables. The technical importance of this in terms of the one-halo term in \wfcode~is discussed in Appendix~\ref{sec:code}. Note that I do not solve the excursion set problem for the modified barrier $\mathcal{G}_X(M)$ but simply substitute the functional form into $\nu$ in the Sheth-Tormen function.

Mass dependence of the barrier is expected due to \emph{scale-dependent growth} present in WDM and FDM. In WDM this is caused by the thermal velocities in the distribution function, which suppress the growth of perturbations on small scales. In FDM this is caused by the scalar gradient energy, or ``quantum pressure,'' which similarly suppresses the growth of perturbations in the full non-linear scalar field equations of motion.

\cite{2001ApJ...558..482B} performed spherical collapse simulations of WDM, where the thermal velocities were modelled using an effective pressure in a fluid description. These simulations gave results for the mass dependence of $\dcrit$, for which \cite{2013MNRAS.428.1774B} provided a fitting function: 
\be
\mathcal{G}_{\rm W}(M) = h_{\rm W}(x)\frac{0.04}{\exp (2.3 x)}+[1-h_{\rm W}(x)]\exp \left[\frac{0.31387}{\exp (0.809 x)} \right] \, , 
\ee
where the auxiliary fitting functions are given by:\footnote{Note I have algebraically simplified the formulae of \cite{2013MNRAS.428.1774B} to remove dependence on $z_{\rm eq}$.} 
\begin{align}
x &= \log (M/M_J) \, , \\
h_{\rm W}(x) &= 1/[ 1+\exp (10x +24)] \, , \\
M_{J,W} & = 2.81 \times 10^8 \left(\frac{m_X}{1\text{ keV}} \right)^{-4} \nonumber \\
&\left(\frac{\Omega_m h^2}{0.15} \right)^{2}\left(\frac{g_X}{1.5} \right)^{-1}\left(\frac{h}{0.7}\right)h^{-1} M_\odot \,  .
\end{align}
The parameter $g_X$ is the number of degrees of freedom, and is given by $g_X=1.5$ for a spin-$\frac{1}{2}$ fermion. It is a free parameter in \wfcode.

$M_{J,W}$ is the WDM Jeans mass, and is the most important parameter in the WDM barrier fit. For $M\ll M_J$, $\mathcal{G}_{\rm W}$ grows as an inverse power of $M$, suppressing the HMF compard to CDM. For $M\gg M_J$, $\mathcal{G}_{\rm W}$ goes to unity exponentially, returning to CDM-like behaviour.

For the FDM barrier I use the model of \cite{2014MNRAS.437.2652M}. This model is based on the physical intuition that if redshift dependence of $\delta_{\rm crit}$ in the Press-Schechter formalism is given by $D(z)$, then in the case of scale-dependent growth, the mass dependence can be modelled using $D(z,k)$. I compute the appropriate growth ratio using results from linear perturbation theory from \textsc{axionCAMB} \citep{2015PhRvD..91j3512H}. The growth ratio oscillates, leading to numerically sensitive features in the HMF, so I smooth the resulting function using a spline. This smoothing is purely cosmetic: the HMF in the one-halo term is integrated over mass and such features are smoothed out and do not affect $\Delta_m^2(k)$.\footnote{Such a smoothing was also done in \cite{2015MNRAS.450..209B}, where again the final results depended on an integral of the HMF and so were insensitive to the barrier smoothing.} 

A fit to the FDM mass-dependent barrier is given by:
\be
\mathcal{G}_{\rm F}(M) = h_{\rm F}(x) \exp [a_3 x^{-a_4}]+[1-h_{\rm F}(x)]\exp [a_5 x^{-a_6}]  \, ,
\ee
where the auxiliary fitting functions are
\begin{align}
x&=M/M_J \, , \\
h_{\rm F}(x) &=(1/2)\{1-\tanh [M_J(x-a_2)]\} \, . \\
M_{J,F}&=a_1\times 10^8 \left(\frac{m_a}{10^{-22}\text{ eV}} \right)^{-3/2}\left(\frac{\Omega_m h^2}{0.14} \right)^{1/4} h^{-1}M_{\odot}\, , \\
\{a_1,&a_2,a_3,a_4,a_5,a_6\}=\{3.4,1.0,1.8,0.5,1.7,0.9\}\, .
\end{align}
The parameters $a_i$ were fit by least squares minimization, and were verified to be only weakly dependent on cosmology. This fit is a centrally useful result of the present work for calculating the HMF for FDM, though as we will see it has little effect on the halo model power spectrum. 

The FDM barrier interpolates between two exponentials around the scale of $M_{J,F}$. It is thus a sharper barrier than the WDM barrier, consistent with the sharper cut-off in the linear theory transfer function, and consistent with physical expectations based on the effect of quantum pressure compared to thermal velocities.

The scalings of the FDM Jeans mass with $\Omega_m$ and $m_a$ were fixed by the scaling of the linear theory Jeans scale (though they were verified in the numerical fit). Note the softer scaling of $M_{J,F}$ with $m_a$ compared to the scaling of $M_{J,W}$ with $m_X$. This is also seen in the linear theory $T_X(k)$ cut-offs.

The WDM and FDM fits to $\mathcal{G}_X(M)$ used in \wfcode~aim to only capture the true behaviours for $\mathcal{G}_X(M)\lesssim 20 \rightarrow 10^2$ at relatively large masses compared to the HMF cut-off. Since the barrier appears in an exponential in the HMF, the cut-off induced by an increase in $\mathcal{G}_X(M)$ is so dramatic that its precise value far from the cut-off is unimportant. For WDM, this was noted also by \cite{2013MNRAS.428.1774B}. This requirement makes the redshift dependence of $\mathcal{G}_{\rm F}(M)$, as calculated from the growth at $z\lesssim 15$, unimportant for $m_a$ allowed by e.g. the CMB \citep{2015PhRvD..91j3512H}. As shown below, the modified barriers, while important for the HMF at $M<M_J$ \citep[leading to the sharp cut-offs seen in e.g.][]{2013MNRAS.428.1774B,2014MNRAS.437.2652M}, they have relatively little effect on the power spectrum on observable scales. 

\subsection{Concentration-mass relationship}
\label{sec:concentrations}

\hmcode~uses the CDM concentration-mass relationship, $c(M)$, for NFW halo density profiles of \cite{2001MNRAS.321..559B}, given here in Eq.~\eqref{eqn:concentration}. \cite{2012MNRAS.424..684S} found, from $N$-body simulations of WDM, the fitting formula:
\be
\frac{c_{\rm WDM}(M)}{c_{\rm CDM}(M)} = \left( 1+\gamma_1\frac{M_{1/2}}{M} \right)^{-\gamma_2} \, ,
\label{eqn:wdm_conc}
\ee
where the fitting parameters are $\gamma_1=15$, $\gamma_2=0.3$, and $M_{1/2}$ is the ``half-mode mass:'' 
\be
M_{1/2} = \frac{4}{3}\pi \bar{\rho}\left( \frac{\pi}{k_{1/2}} \right)^3
\ee
defined from the wavenumber $k_{1/2}$, such that $T_{\rm W}(k_{1/2})=0.5$. 

Note that in order to use the fitting formula Eq.~\eqref{eqn:wdm_conc}, $c_{\rm CDM}(M)$ in \wfcode~is computed using the CDM linear power. This accounts for an observed \emph{increase} in the halo concentration for WDM halos compared to the expectation using the WDM linear variance and the model of \cite{2001MNRAS.321..559B} (the concentration is still reduced compared to CDM). The same should conceivably be true for FDM, given the qualitative similarities to WDM \citep[e.g. $N$-body simulations of][]{2016ApJ...818...89S,2016JCAP...04..012S}. In \wfcode~I use Eq.~\eqref{eqn:wdm_conc} also for FDM, with $k_{1/2,{\rm F}}=0.5 k_{J,{\rm eq}}m_{22}^{-1/18}$~\citep{2000PhRvL..85.1158H}. 

\cite{2012MNRAS.424..684S} observed that the most important effect of using a different $c(M)$ was on the shape of the power near the WDM cut-off. This leads to the power at $z=0$ being increased by around 10\% in the range $1\lesssim [k/(h\text{ Mpc}^{-1})]\lesssim 10$ compared to using the \cite{2001MNRAS.321..559B} $c(M)$. The increase in power, arising from the increase in halo concentration, leads to better agreement with $N$-body simulations. 

\subsection{The Non-linear Power Beyond CDM}
\label{sec:examples}

\begin{figure*}
\includegraphics[width=2\columnwidth]{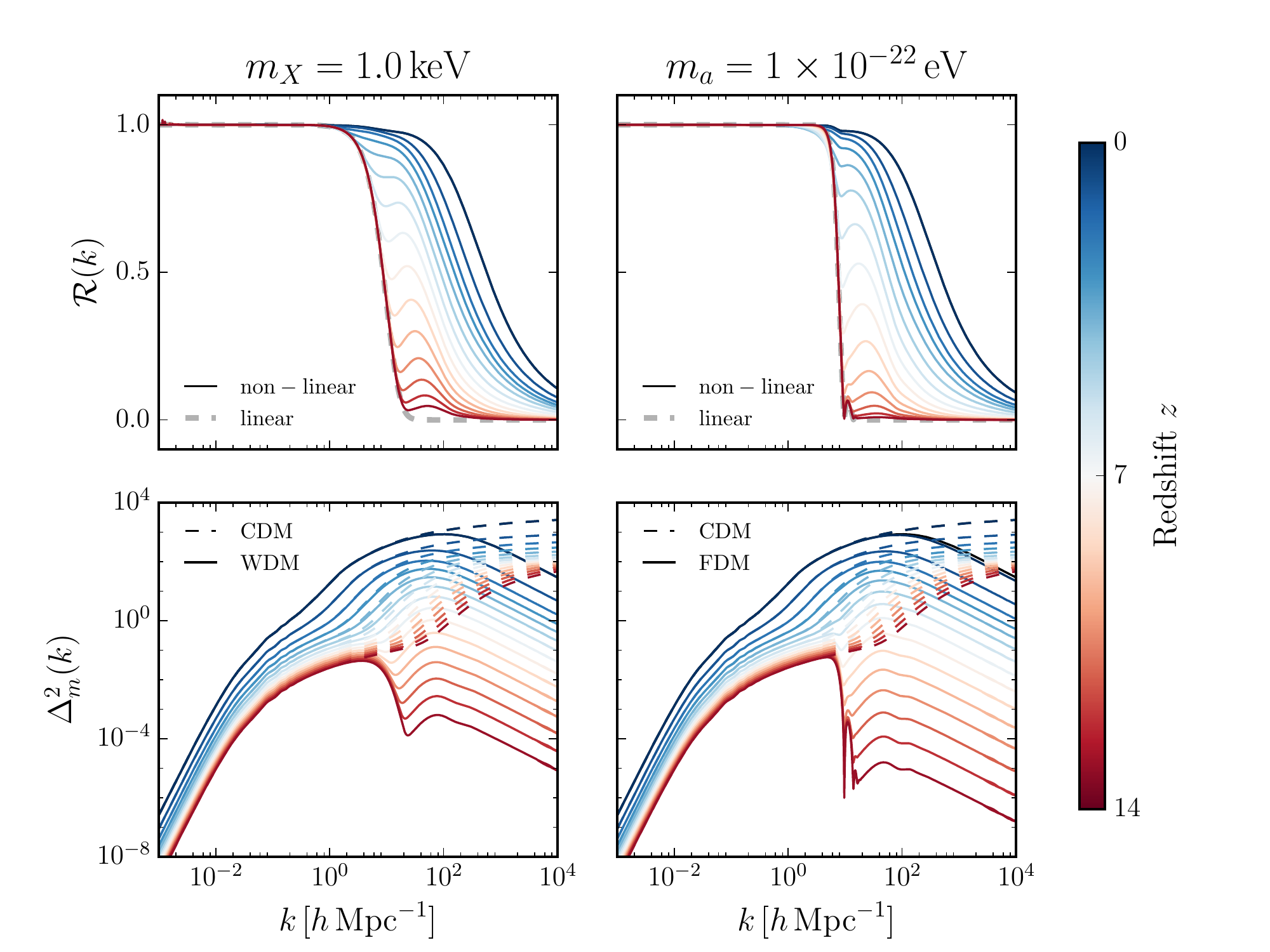}
\caption{Power ratios (Eq.~\ref{eqn:calR_def}) and power spectra for WDM and FDM computed using \wfcode. The power ratios are compared to the linear theory (the transfer function), which is a good description at high-$z$. The power spectra are compared to CDM, which is a good description at low-$k$. Non-linear growth reduces the amount of power suppression relative to CDM at intermediate-$k$ compared to linear theory. However, both WDM and FDM still display a marked suppression of power at high-$k$, and shape differences between the two models at intermediate-$k$. }
\label{fig:tk_dk_m1}
\end{figure*}	

Fig.~\ref{fig:tk_dk_m1} shows the non-linear (NL) power spectra, $\Delta_m^2(k)$, and power ratios relative to CDM:
\be
\mathcal{R}_{X}(k)=\sqrt{\Delta^2_X(k)/\Delta^2_{\rm CDM}(k)}\, .
\label{eqn:calR_def}
\ee
in the redshift range $z\in [0,14]$ for the benchmark models with $m_X=1\text{ keV}$ and $m_a=10^{-22}\text{ eV}$, computed using \wfcode.\footnote{The power is shown out to $k=10^4\,h\text{ Mpc}^{-1}$, however baryonic feedback is expected to become important at much lower wavenumbers. Thus, the spectra shown are a representation of the DM-only effects. Feedback can be included following e.g. the analytic models of M15 and \cite{2015MNRAS.451.2877M}, after matching to simulations.} In linear theory, we have that $\mathcal{R}(k)$ is equal to the transfer function, $T(k)$.

At high-$z$, halo formation is greatly suppressed in both WDM and FDM compared to CDM, and linear theory provides a good description of $\mathcal{R}(k)$ near the cut-off. In the non-linear theory, the cut-off in power for both WDM and FDM is moved to larger wavenumbers (smaller scales), with this effect being more pronounced at low-$z$. At $z=0$ the cut-off in $k$ in both models is increased by approximately two orders of magnitude compared to linear theory.

Non-linear collapse removes some of the differences between WDM and FDM in the deeply non-linear regime. This is because the one-halo term, Eq.~\eqref{eqn:oneH_term}, is an integrated quantity, erasing detailed dependence on the shape of the linear theory transfer function and the collapse barrier. However, there are still clear differences in the shape of $\mathcal{R}_X(k)$ for WDM and FDM near the onset of the cut-off. This has significant implications for constraints to FDM from the Lyman-$\alpha$ forest flux power spectrum, which will be discussed in a companion paper. 

\begin{figure*}
\includegraphics[width=2\columnwidth]{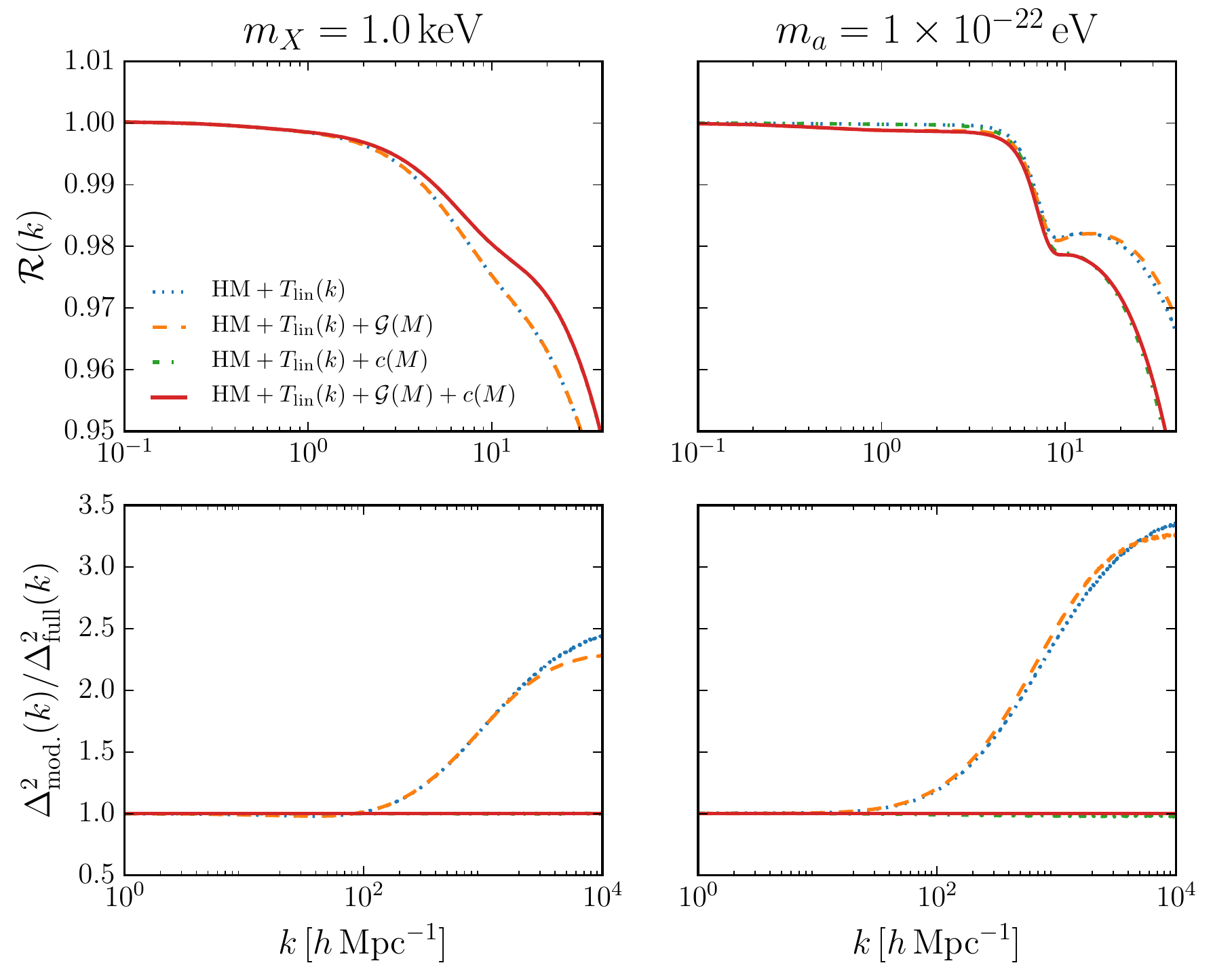}
\caption{Effects of different aspects of the halo model (HM) at $z=0$. Note the different $x$-scales between the top and bottom row. \emph{Top row:} Power ratios (Eq.~\ref{eqn:calR_def}) relative to CDM. \emph{Bottom row:} Relative effects of each piece of the halo model compared to the full model. The largest effect by far, after the linear transfer function, is given by the modified $c(M)$, which overall suppresses power at $k\gtrsim 10^2\,h\text{ Mpc}^{-1}$. However, as evidenced in the top left panel, the modified $c(M)$ increasea power for WDM over a range of intermediate scales, consistent with \citep{2012MNRAS.424..684S}. This does not occur for FDM, as evidenced in the right hand column, where the modified $c(M)$ relation leads to suppression of power on all scales.}
\label{fig:conc_barrier_4panel}
\end{figure*}	

Fig.~\ref{fig:conc_barrier_4panel} explores the effects of different aspects of the halo model employed in \wfcode~on both WDM and FDM, at $z=0$. The top left panel shows how the modified $c(M)$ leads to an increase in power for WDM over a range of scales. The top right panel shows that this is not the case for FDM, where each new effect included in the model further suppresses power. For both FDM and WDM, the overall effect of $c(M)$ is to suppress power on small scales, where it becomes by far the dominant effect. The modified barrier, $\mathcal{G}(M)$ also suppresses power at the smallest scales in both models. The effect of $\mathcal{G}(M)$, while important in the HMF, is relatively unimportant in the power spectrum.

\section{Discussion}
\label{sec:discussion} 

\wfcode~presents a simple, computationally fast, version of the halo model for FDM and WDM. It is built on the form of the halo model used in \cite{2015MNRAS.454.1958M} and given in Appendix~\ref{sec:halomodel}. Compared to the more comprehensive halo model of WDM by \cite{2011PhRvD..84f3507S,2012MNRAS.424..684S}, \citep[see also][]{2011arXiv1109.6291D}, \wfcode~makes a number of approximations:
\begin{itemize}
\item Approximate two-halo term.
\item Concentration relation assumed as for WDM.
\item Clustering of the smooth component ignored.
\item Halo density profiles assumed as for CDM.
\item FDM with no self-interactions.
\end{itemize}

I now address the expected importance of each of these points in turn. All items on this list should strictly be investigated in dedicated simulations, as was done for WDM in, e.g., \cite{2012MNRAS.424..684S}. Accounting for all these effects, the halo model for WDM was found to give an absolute accuracy of $\sim 10\%$ compared to simulations, and was able to predict the \emph{relative} effect of WDM compared to CDM to within $\sim 5\%$, with the accuracy improving at high-$k$. There are no such simulations for FDM, so the accuracy of the \wfcode~for FDM is hard to assess quantitatively. 

In addition, it should be noted that, as with ordinary $N$-body and related codes,\footnote{For a recent exception, see \cite{2016arXiv160406065A}.} and applications of the halo model and cosmological perturbation theory, \wfcode~only applies in the non-relativistic limits of sub-horizon scales, small curvature, and, in the case of FDM, time and length scales long compared to $1/m_a$.

\emph{Two-halo term:} \hmcode~approximates the two-halo term as proportional to the linear power spectrum, with an additional model for the damping in the quasi-linear regime fit to CDM simulations and emulators (see Eq.~\ref{eqn:mead_quasilin_mod}). The full two-halo term is given by
\be
P_{\rm 2H} = P_{\rm lin}(k)\prod_{i=1,2}\int dM_i b(M_i)\frac{M_i}{\bar{\rho}}W(k,M_i)F(M_i) \, .
\label{eqn:full_twoh}
\ee
The functions $F(M_i)$ and $W(k,M_i)$ are, respectively, the HMF and the Fourier-transformed halo density profile, as defined in Appendix~\ref{sec:halomodel}. In Eq.~\eqref{eqn:full_twoh} the bias in the halo-halo power, $P_{\rm hh}(k)$, has been taken to be
\be
P_{\rm hh}(k|M_1,M_2)\approx b(M_1)b(M_2)P_{\rm lin}(k) \, ,
\ee
where $M_1$ and $M_2$ are the halo masses. Given such a model, the bias can be fit from simulations. \cite{2011PhRvD..84f3507S} modelled the WDM bias to be the same for CDM and WDM. \cite{2012MNRAS.424..684S} observed that, for halo masses above the HMF cut-off, the WDM bias was well described by the CDM case. 

Thus, relative to the approximation used in \wfcode, the full two-halo term picks up additional $k$-dependence from the integrated effect of the HMF, density profiles, and bias. Without switching to the full two-halo term within \wfcode~none of these effects can yet be modelled. The most important effect is likely due to $c(M)$ in the quasi-linear regime, which I now discuss.

\emph{Concentration-mass relation:} \wfcode~assumes the WDM concentration-mass relationship, $c(M)$, of \cite{2001MNRAS.321..559B,2012MNRAS.424..684S}. This is assumed to be the same for WDM and FDM, depending only on the half-mode in each model, which is clearly an approximation that should be checked with simulation of FDM. 

As mentioned above, the primary effect of modifying $c(M)$ for WDM was to give a boost in the power on intermediate scales, giving better agreement with simulation. In \wfcode~a boost in power is seen using the modified $c(M)$. However, the boost does not kick in until larger $k$ compared to the model used in \cite{2012MNRAS.424..684S}, and is less pronounced (c.f. Fig.~\ref{fig:conc_barrier_4panel}, top left panel in this paper, and their Figs. 13 and 15). This is because part of the boost in power is due to the effect of $c(M)$ in the full two-halo term, which is absent in \wfcode, suggesting there is an error of around 10\% in the absolute value of the power for $1\lesssim [k/(h\text{ Mpc}^{-1})]\lesssim 10$.

The use of the WDM $c(M)$, and the absence of $c(M)$ effects in the approximate two-halo term, are thus expected to be the major source of error in \wfcode~compared to the ``true'' power in simulations.

\emph{Smooth component:} In WDM and FDM models, due to the cut-off in the power, only a fraction, $f_h$, of the total DM mass is contained in halos, and there can be a large smooth component. The smooth component of DM has linear clustering, cross correlation with the DM in halos, and its own bias. The smooth component modifies the total power to be
\begin{align}
\Delta_m^2(k)&=(1-f_h)^2 \Delta_{\rm ss}^2(k)+2f_h(1-f_h)\Delta_{\rm sh}^2(k) \nonumber \\
&+f_h^2 [\oneh+\twoh]\, .
\end{align}
The formulae defining $\Delta_{\rm ss}^2(k)$ and $\Delta_{\rm sh}^2(k)$ in terms of the linear power, HMF, and bias can be found in \cite{2011PhRvD..84f3507S,2012MNRAS.424..684S}. Both depend on the linear power, and thus become sub-dominant to the one-halo term at large-$k$. 

The fraction of DM in halos, $f_h$, is given by
\be
f_h=\frac{1}{\bar{\rho}}\int_{0}^\infty dM M \frac{dn}{dM} \, ,
\ee
where $f_h$ is defined to be unity in a universe with perfectly hierarchical structure formation (i.e. for CDM). The cut-off in the HMF for WDM and FDM leads to $f_h<1$: the clustering of the smooth components contributes to the total power, and the importance of the two and one-halo terms in partially suppressed.

In the WDM halo models of \cite{2011PhRvD..84f3507S,2011arXiv1109.6291D,2012MNRAS.424..684S} the power spectrum was shown always at $z=0$ (with the exception of the lensing power, which is an integral) and it was generally concluded that the smooth component has a minor effect, since $f_h$ is close to unity at low redshift for the models considered. 

However, the cut-off in the HMF leads to dramatic suppression of halo formation at high-$z$, with for example, no halos at all expected at $z\gtrsim 10$ and $m_a\approx 10^{-22}\text{ eV}$~\citep{2015MNRAS.450..209B,2016ApJ...818...89S}. Thus, at high-$z$, $f_h\rightarrow 0$ and the power should return to essentially linear (in particular, no one-halo term). This will significantly affect the shape of the absolute power spectrum at high $z$ and $k$ shown in the bottom row of Fig.~\ref{fig:tk_dk_m1}. However, this will have very little effect on the power ratio, since the suppression of the one-halo term is already large enough to send $\mathcal{R}(k)$ essentially to zero on a linear scale.

Interestingly, a prescription to send the power to linear when the typical fluctuation is small, $\sigma<1$, is built in to the form of \textsc{halofit}~\citep{2003MNRAS.341.1311S} used in \textsc{camb}. While this is done for computational simplicity in models of CDM, it has some relation to the physical role of $f_h$ for FDM and WDM. One possible observational consequence is for the CMB lensing power spectrum in mixed DM models with low $m_{a/X}$, where $f_h$ can be close to zero near the peak of the lensing kernel at lower $z\approx 2$. Another possible consequence is for the 21cm power in the dark ages.

In summary, \wfcode~does not model the smooth component. This implies that the power at high-$z$ is expected to be much closer to the linear power, with no one-halo term. Including the smooth component in \wfcode~is left for future work.

\emph{Halo density profiles:} The halo model in \wfcode~employs the NFW halo profile, Eq.~\eqref{eqn:nfw_profile}. WDM and FDM halos are expected to deviate from the NFW profile. This is partially captured in CDM-like $N$-body simulations, where profiles may be slightly flattened due to the different formation history caused by the truncated initial power \citep[e.g][]{2011arXiv1109.6291D,2000ApJ...542..622C,2001ApJ...559..516A}, though the importance of this effect depends on simulation resolution \citep[e.g.][]{2012MNRAS.424..684S}.

However, larger effects occur on smaller scales, where WDM and FDM deviate from standard $N$-body treatments. These effects lead to the formation of density cores on small scales $r_{\rm core}\lesssim 1\text{ kpc}$. 

For WDM the cores are formed due to the thermal velocities and fine grained phase space \citep{1979PhRvL..42..407T}. By computing the phase space density, the core size in a halo can then be calculated, and added to the halo model \citep[as in][]{2011PhRvD..84f3507S} or to simulations \citep{2012MNRAS.424.1105M,2013MNRAS.428.3715M,2013MNRAS.430.2346S}.

For FDM, cores are formed due to the de Broglie wavelength in the underlying scalar wave equation (the so-called ``quantum pressure''). This leads to the formation of stable solitonic cores in halos~\citep[e.g.][]{1969PhRv..187.1767R,1991PhRvL..66.1659S}, which are found in full cosmological simulations~\citep{2014NatPh..10..496S}. Such cores can be modelled analytically in a variety of ways \citep[e.g.][]{2014MNRAS.437.2652M,2015MNRAS.451.2479M}. 

Deviations of the WDM and FDM density profiles from CDM are not expected to have large effects on the power spectrum computed by \wfcode~on observable scales. This was shown to be true for WDM by \cite{2011PhRvD..84f3507S}, where cores were estimated to affect the power spectrum by less than 1\% for wavenumbers $k<100 \, h \text{ Mpc}^{-1}$ and $m_X>0.25\text{ keV}$. For FDM, the core-halo mass relationship leads to larger cores than for WDM with a similar power spectrum cut-off~\citep{2014PhRvL.113z1302S,2014MNRAS.437.2652M,2015MNRAS.451.2479M}, so the effect of cores on the power spectrum could be larger. Incorporating modifications to the simple NFW profile into \wfcode~is left as a topic for future work, but for the power spectrum on observable scales the effects are expected to be small.

\emph{FDM self-interactions:} \wfcode~essentially assumes that FDM is described by the scalar potential $V(\phi)=m_a^2\phi^2/2$, with the field coherently oscillating about the potential minimum. The results for the power spectrum should not depend too drastically on this assumption: any field coherently oscillating in a quadratic minimum will behave as matter on large scales~\citep{1983PhRvD..28.1243T}, and experience acoustic oscillations on small scales~\citep{khlopov_scalar}. Thus, the large-field behaviour of the potential is not expected to affect the linear transfer function much. 

Self-interactions may have larger effects on spherical collapse, changing the shape of $\mathcal{G}(M)$. As already noted, the exact shape of $\mathcal{G}(M)$ does not have much effect on the halo model, and in addition \wfcode~fits to the linear growth in $m_a^2\phi^2$, and so self-interactions are not expected to be important here either.

FDM self interactions can have large effects on the halo density profile, for example distinguishing between axions, with attractive interactions, and other scalars, with repulsive interactions~\citep[e.g.][]{2011PhRvD..84d3531C,2011PhRvD..84d3532C,2015PhRvD..92j3513G}. However, as already noted, halo density profiles beyond NFW are not modelled in \wfcode~and are expected to have only small effects on the power spectrum. Extending \wfcode~beyond $m_a^2\phi^2$ is left for future work. Qualitative results of \wfcode~at high-$k$, and quantitative results in the quasi-linear regime, are expected to apply to all models of scalar field/axion DM with matter-like oscillations about a quadratic potential minimum.

\section{Conclusions}
\label{sec:conclusions}

In conclusion, \wfcode~is a simple and fast way to compute the non-linear power spectrum using the halo model for the popular WDM and FDM models. The computation is based on \hmcode~by \cite{2015MNRAS.454.1958M}, though \wfcode~does not by default use the additional tunings of \hmcode. \wfcode~ is expected to be quantitatively correct at the $\mathcal{O}(10\%)$ level for $k\lesssim 10^2\, h\text{ Mpc}^{-1}$, and possibly on smaller scales. Results at very high $k$ are indicative of general behaviour only. Future work will involve testing \wfcode~against the appropriate simulations ($N$-body or otherwise), and refining the model further. Particular refinements will be to improve the two-halo term and concentration mass relationship, which will improve accuracy in the quasi-linear regime, and to include modified halo density profiles, for physical completeness. Including the clustering of the smooth component of FDM and WDM will have considerable effects on the power at large $z\gtrsim 8$.

Companion papers are in preparation, which will apply the results of \wfcode~to the Lyman-$\alpha$ forest flux power spectrum, the CMB lensing power spectrum, and other observables.

\wfcode~is publicly available at \url{https://github.com/DoddyPhysics/HMcode}, and collaboration on development is welcomed.

\section*{Acknowledgments}

I acknowledge many useful conversations with Alexander Mead on the development of \textsc{WarmAndFuzzy} from \textsc{HMcode}, and for comments on the draft. I also acknowledge useful conversations with Daniel Grin. I am supported by a Royal Astronomical Society Fellowship hosted at King's College London.

\bibliography{WarmAndFuzzy}
\bibliographystyle{mnras}

\appendix

\section{The Halo Model}
\label{sec:halomodel}

I present here the halo model as used in \hmcode~and \wfcode, following M15, but remind the reader that the additions in M15 (made clear below) are not used by default.

The matter power spectrum, $P(k)$, is the the Fourier transform of the two-point correlation function of the matter density fluctuation, $\delta (x)$. We work with the dimensionless power:
\be
\Delta_m^2(k) = 4\pi \left(\frac{k}{2\pi}\right)^3 P(k) \, .
\ee

The first key ingredient in the halo model is the variance of the linear power spectrum at $z=0$, $\sigma^2 (R)$, which we define as:
\be
\sigma^2(R) = \int_0^\infty d\ln k \Delta_{\rm lin}^2(k) \mathcal{W}(kR) \, ,
\ee 
where the window function, $\mathcal{W}(kR)$, is the Fourier-transform of a spherical top-hat:
\be
\mathcal{W}(x) = \frac{3}{x^3}(\sin x - x\cos x)\, .
\ee
Using the enclosed mean density, the variance can be mapped to a function of mass, $\sigma^2(M)$.\footnote{The real-space window function allows for unambiguous assignment of the halo mass. Note that some authors advocate a sharp-$k$ window function to capture the shortcomings on Press-Schechter as applied to models with suppressed power, such as FDM and WDM. For further discussion on these issues, see e.g. \cite{2013MNRAS.433.1573S}, and for FDM, \cite{2015arXiv151108195U}.}

The halo model expresses the power as the sum of a `one-halo' and a `two-halo' term:
\be
\Delta_m^2(k) = \oneh+\twoh \, ,
\ee
where the one-halo term, \oneh, represents shot noise from a random distribution of halos with known mass function and density profiles, and the two-halo term, \twoh, represents two-point correlations between the halos. This relationship is modified in M15 to better model the transition between the quasi-linear and fully non-linear regime:
\be
\Delta_m^2(k) = [(\oneh)^\alpha+(\twoh)^\alpha]^{1/\alpha} \, ,
\label{eqn:mead_quasilin_mod}
\ee
where $\alpha$ is fit from simulations as $\alpha=2.93\times 1.77^{n_{\rm eff}}$, with $n_{\rm eff}$ the effective index of the linear power spectrum variance at the non-linear scale:
\be
3+n_{\rm eff} = -\left|\frac{d \ln \sigma^2(R)}{d\ln R}\right|_{\sigma=1} \, .
\ee

In the simplest case, the two-halo term can be approximated by the linear-theory power:
\be
\twoh = \Delta^2_{\rm lin}(k) \, .
\ee
I discussed the linear theory power and growth for CDM, FDM, and WDM, in Section~\ref{sec:main}. In Section~\ref{sec:discussion}, I discussed the effects of approximating the two-halo term this way.

Linear theory over predicts the $z=0$ power on quasi-linear scales. M15 modifies the two-halo term with the inclusion of an additional damping term. The model for the damping is predicted from perturbation theory~\citep{2006MNRAS.373..369C}, and trucated at quadratic order to best-fit simulations:
\be
\twoh = \left[1-f_{\rm 2H}\tanh^2(k \sigma_V/\sqrt{f_{\rm 2H}}) \right]\Delta^2_{\rm lin}(k) \, .
\label{eqn:mead_2H_mod}
\ee 
The damping scale is the linear displacement variance
\be
\sigma_V^2 = \frac{1}{3}\int_0^\infty dk \frac{\Delta^2_{\rm lin}(k)}{k^3}\, .
\ee
The parameter $f_{\rm 2H}$ is fit in M15 as $f_{\rm 2H}=0.188[\sigma_8(z)]^{4.29}$, where $\sigma_8$ is the linear power spectrum normalisation on $8 \,h^{-1}\text{ Mpc}$ scales, as usually defined.

The one-halo term is given by the convolution in halo-mass space of the HMF with the halo density profile:
\be
\oneh = 4\pi \left(\frac{k}{2\pi}\right)^3\frac{1}{\bar{\rho}^2}\int_0^\infty dM M^2 W^2(k,M)F(M) \, ,
\label{eqn:oneH_term}
\ee
where $\bar{\rho}=\Omega_m\rho_{\rm crit}=3\Omega_m H_0^2/(8\pi G)$ is the mean cosmic matter density. The window function, $W(k,M)$, is defined by the normalised Fourier transform of the halo density profile, $\rho_{\rm halo}(r,M)$:
\be
W(k,M) = \frac{1}{M}\int_0^{r_v}dr \frac{\sin (kr)}{kr}4\pi r^2 \rho_{\rm halo}(r,M) \, ,
\ee
where $r_v$ is the halo virial radius (defined in this context below). 

The HMF, $F(M)\equiv dn/dm$, with $dn$ the comoving halo number density, is defined by \citep{1974ApJ...187..425P}:
\be
\frac{dn}{dM}dM =\frac{\bar{\rho}}{M}  f(\nu) \frac{d\nu^2}{\nu^2} \, .
\label{eqn:hmf_def_standard}
\ee  
In the excursion set formalism \citep{1991ApJ...379..440B} the function $f(\nu)$ is the first crossing statistic for random walks, which is universal for a given barrier shape. The variable $\nu$ expresses the ratio between the variance of matter fluctuations on scale $M$, $\sigma(M)$, to the ``critical barrier for collapse,'' $\delta_{\rm crit}$:
\be
\nu \equiv \frac{\delta_{\rm crit}}{\sigma (M)} \, .
\ee
In the standard halo model, $\sigma(M)$ is taken at redshift $z=0$, and redshift evolution is included in the barrier, which is assumed mass-independent. I take the universal mass function, $f(\nu)$, from \cite{1999MNRAS.308..119S}:
\be
f(\nu) = A\sqrt{\frac{a}{2\pi}}\nu [1+(a\nu^2)^{-p}]\exp (-a\nu^2/2) \, ,
\label{eqn:st_mass_function}
\ee
with $A=0.322$, $a=0.707$, $p=0.3$. Section~\ref{sec:barriers} discusses how I modify the mass function in the case of a mass-dependent barrier. Note that Eq.~\eqref{eqn:st_mass_function} can be derived for an ellipsoidal barrier \citep{2001MNRAS.323....1S}.

In \hmcode~$\delta_{\rm crit}^0$ is given an additional, mild, cosmology and redshift dependence in the fit to simulations for CDM:
\be
\dcrit^0 = 1.59+0.0314 \ln \sigma_8 (z) \, .
\label{eqn:mead_dc_mod}
\ee

M15 modifies the one-halo term with a damping-term to prevent it dominating over the linear power on largest scales, and an additional parameter $\eta>0 (<0)$ that increases (decreases) power due to ``puffed up'' higher-mass halos:
\begin{align}
\oneh = &[1-\exp (-k/k_\star)^2] \, \nonumber \\ 
& 4\pi \left(\frac{k}{2\pi}\right)^3\frac{1}{\bar{\rho}^2}\int_0^\infty dM M^2 W^2(\nu^\eta k,M)F(M) \, .
\label{eqn:mead_1H_mod}
\end{align}
The damping scale, $k_\star=0.584 \sigma_V^{-1}(z)$, and $\eta = 0.603-0.3\sigma_8(z)$. 

The final ingredient in the halo model is the halo density profile, $\rho_{\rm halo}(r,M)$. In \wfcode, I treat WDM and FDM halos exactly as CDM halos (see Section~\ref{sec:discussion} for discussion on this approximation), with the halo density profile given by the usual NFW profile:
\be
\rho_{\rm halo}(r,M) = \frac{\rho_{\rm N}}{(r/r_s)(1+r/r_s)^2} \, ,
\label{eqn:nfw_profile}
\ee
where $r_s$ is the scale radius of the halo and $\rho_N$ is a normalisation used to fix the halo mass, $M$, via the spherical integral of the density profile. The mass is fixed using an overdensity threshold, $\Delta_V$, at which the profile is truncated, and which defines the virial radius, $r_V$ from the enclosed mean density:
\be
M = \frac{4}{3}\pi\bar{\rho} \Delta_V  r_V^3 \, .
\ee 
The limitations of using the NFW profile to describe FDM and WDM in the halo model are discussed in Section~\ref{sec:discussion}.

The halo concentration, $c$, is defined as the ratio of the virial radius to the scale radius of the halo, $c\equiv r_V/r_s$. Therefore, once an overdensity threshold, $\Delta_V$, has been set, there is a single free function that defines all NFW halos: the concentration-mass relationship, $c(M)$. \wfcode~sets $\Delta_V=200$. M15 uses the fit:
\be
\Delta_V = 418\times \Omega_m(z)^{-0.352}\, .
\label{eqn:mead_virial_mod}
\ee

The concentration-mass relationship used in M15 is taken from \cite{2001MNRAS.321..559B} and \cite{2004A&A...416..853D}:
\be
c(M,z) = A \frac{1+z_f(M)}{1+z}\frac{D_{\rm DE}(z\rightarrow \infty)}{D_{\Lambda}(z\rightarrow \infty)} \, ,
\label{eqn:concentration}
\ee
where $A=3.13$ is normalised from simulation, $z_f(M)$ is the formation redshift of a halo with mass $M$, and $D_X(z)$ is the usual linear theory growth function in cosmology $X$ normalised such that $D_X(0)=1$.\footnote{The growth function correction in Eq.~\eqref{eqn:concentration} only applies when $w\neq -1$, where $w$ is the dark energy (DE) equation of state (taking the standard $(w_0,w_a)$ parameterisation in \hmcode). In the interests of ``one tooth-fairy at a time,'' I do not advise combining $w\neq -1$ cosmologies with FDM and WDM, not least since there are not linear transfer functions, never mind simulations, to calibrate these combined departures from $\Lambda$CDM.} Variation of the growth function in FDM and WDM is absorbed in the barrier function, as discussed in Section~\ref{sec:barriers}.

The halo formation redshift is defined by the redshift at which a fraction $f_{\rm coll}$ of mass has collapsed into the halo. \cite{2001MNRAS.321..559B} simplify the NFW relation with a model that better fits simulations:\footnote{Note that I have cancelled one factor of the linear growth relative to that in M15. Note also that the FDM halo model in \cite{2014MNRAS.437.2652M} used the NFW formula, based on \cite{1993MNRAS.262..627L}.}
\be
D(z_f)\sigma (f_{\rm coll}M)=\delta_{\rm crit} \, .
\label{eqn:collapse_redshift}
\ee
When Eq.~\eqref{eqn:collapse_redshift} is inverted to find $z_f$, if it is found that $z_f<z$, i.e. halo formation ``in the future,'' then \hmcode~sets $c=A$ above. The value $f_{\rm coll}=0.01$ is fit from simulations.

The $c(M,z)$ relationship used in \wfcode~uses the fitting formula Eq.~\eqref{eqn:wdm_conc} of \cite{2012MNRAS.424..684S}, which uses the CDM linear variance as input, and \emph{not} the linear variance for WDM or FDM. 

\section{Modifications to \hmcode}
\label{sec:code} 

\wfcode~is a modification of \hmcode~(presented in M15) and is written in \textsc{Fortran90}. It should compile with most standard compilers, and does not need to be pointed to any libraries. I compile with \textsc{gfortran}. Most major modifications in \wfcode~are commented beginning ``WFcode.''

Two binary parameters, \textsc{ifdm}$=0,1$, and  \textsc{iwdm}$=0,1$, are used to turn on and off FDM and WDM treatments respectively. They cannot be used together, so at least one of these parameters must be set to zero. 

The FDM mass is given in units of $10^{-22}\text{ eV}$, and the WDM mass is given in units of keV. \wfcode~is not tested or expected to be reliable for very small values of these parameters, where FDM or WDM start to affect matter-radiation equality. The code will stop if either mass is less than $10^{-2}$ in the given units. In any case, masses lower than this for the dominant DM component are inconsistent with observations of e.g. the CMB.

As discussed in Section~\ref{sec:main}, FDM and WDM are implemented in \wfcode~with three separate pieces: modifications to the linear-theory transfer function, modifications to the collapse barrier, and modifications to the concentration-mass relationship. The modification to the linear transfer function is always used. The other modifcations can be implemented separately, using the parameters \textsc{ibarrier}$=0,1$ and \textsc{iconc}$=0,1$. Setting either parameter to zero turns off that modification. 

With \textsc{iconc}$=0$ the $c(M)$ relationship is that of \cite{2001MNRAS.321..559B} with the appropriate linear variance \citep[for WDM this is the same as the one used by][]{2011PhRvD..84f3507S}. When \textsc{iconc}$=1$, one must compute $c_{\rm CDM}(M)$ in order to use Eq.~\eqref{eqn:wdm_conc}. This requires a number of extra functions and data structures to be set up inside \wfcode~that hold the CDM results. 

\hmcode~and \wfcode~have an in-built minimum halo mass, $M_{\rm min}$. In testing it was found that, with \textsc{ibarrier}$=1$, results become unstable if the minimum halo mass is set to be too small, as integrations need to extrapolate the HMF far below the cut-off. If \textsc{ibarrier}$=1$, the minimum halo mass in \wfcode~ is changed from its default value of $10^{2}h^{-1}M_\odot$ to $10^{-1}\times M_J$, where $M_J$ is the Jeans mass defined by the fit for $\mathcal{G}_X(M)$. This should provide accurate results, since the HMF is cut off strongly at $M_J$, with the minimum not affecting the overall value of the integral.

\hmcode~includes several modifications to the basic halo model, as described above and in Table 2 in M15 (in this paper, Eqs.~\ref{eqn:mead_dc_mod}, \ref{eqn:mead_quasilin_mod}, \ref{eqn:mead_2H_mod}, \ref{eqn:mead_1H_mod}, \ref{eqn:mead_virial_mod}). These are turned on and off with \textsc{imead}$=0,1$. The two modifications of M15 and \wfcode~can be used together, but it should be noted that the parameters of the \textsc{imead} modifications are tuned to simulations with CDM (+massive neutrinos and/or modified gravity). The principles of the basic halo model should be generally applicable to the beyond-CDM models of \wfcode. The modifications for FDM and WDM are physically motivated, but are \emph{not} tuned to non-linear simulation results for $\Delta_m^2(k)$. Such tuning is welcomed as ongoing work or contribution to \wfcode. Therefore, I advise using the setting \textsc{imead}$=0$ for WDM and FDM. In particular, the results shown in Fig.~\ref{fig:tk_dk_m1} were computed with \textsc{imead}$=0$. It was noticed in testing that \textsc{imead}$=1$ gave excess power to WDM/FDM over CDM models near the cut-off at intermediate redshifts, which appears broadly inconsistent with simulations. 

The final technicality of \wfcode~compared to \hmcode~arises form the difference between Eq.~\eqref{eqn:hmf_def_mass_barrier} and Eq.~\eqref{eqn:hmf_def_standard}: with a mass-dependent barrier, $\sigma$ and $\dcrit$ are independent variables. \hmcode~changes variables in the one-halo integral, Eq.~\eqref{eqn:oneH_term}, to integrate over $\nu$. However, according to Eq.~\eqref{eqn:hmf_def_mass_barrier}, the change of variables in the case of a mass-dependent barrier should really be to an integral over $\sigma$. Hence, \wfcode~does the integral over $\sigma$. I verified that the change of variables does not lead to numerical errors in the CDM case (where the integrals are analytically equivalent). Explicitly, the one-halo integral in \wfcode~is given by (without the additional fits of M15)
\be
\oneh = 4\pi \left(\frac{k}{2\pi}\right)^3\frac{1}{\bar{\rho}}\int_{\sigma_{\rm min}}^{\sigma_{\rm max}}d\sigma \, \, M W^2 \tilde{f}(\nu,\sigma) \, ,
\ee
where $\sigma_{\rm min}$ and $\sigma_{\rm max}$ are set by $M_{\rm min}$ and $M_{\rm max}$ and
\be
\tilde{f}(\nu,\sigma)= -\left(\frac{\nu}{\sigma}\right)\tilde{A} [1+(a\nu^2)^{-p}]e^{-a\nu^2/2} \, ,
\ee
with $\tilde{A}=0.2162$, $a=0.707$, $p=0.3$.

\end{document}